\documentclass[aps,preprint,showpacs,showkeys,nofootinbib,floatfix,superscriptaddress]{revtex4}

\usepackage{graphicx}  
\usepackage{bm}  
\usepackage{amsmath}
\newcommand{\beq}{\begin{equation}}
\newcommand{\eeq}{\end{equation}}
\newcommand{\bea}{\begin{eqnarray}}
\newcommand{\eea}{\end{eqnarray}} 
\newcommand{\beqa}{\begin{eqnarray}}
\newcommand{\eeqa}{\end{eqnarray}}

\begin{document}
\title{Scaling functions applied to three-body recombination of $^{133}$Cs  atoms}
\author{L. Platter}\email{lplatter@mps.ohio-state.edu}
\affiliation{Department of Physics,
The Ohio State University, Columbus, OH\ 43210\\}
\affiliation{Department of Physics and Astronomy,
Ohio University, Athens, OH 45701, USA\\}
\author{J.~R.~Shepard}\email{James.Shepard@Colorado.Edu}
\affiliation{Department of Physics, University of Colorado,
Boulder, CO 80309, USA\\}


\date{\today}
\begin{abstract}
We demonstrate the implications of Efimov physics in the
recently measured recombination rate of $^{133}$Cs atoms.
By employing previously calculated results for the energy dependence
of the recombination rate of $^4$He atoms, we obtain three independent
scaling functions that are capable of describing the recombination
rates over a large energy range for identical bosons with
large scattering length. We benchmark these and previously obtained scaling
functions by successfully comparing their predictions with full
atom-dimer phase shift calculations 
with artificial $^4$He potentials yielding large scattering lengths. 
Exploiting universality, we finally use these functions to determine the
3-body recombination rate of $^{133}$Cs atoms with large positive
scattering length, compare our results to experimental data
obtained by the Innsbruck group and find excellent agreement.
\end{abstract}


\maketitle

\section{Introduction}
In atomic physics the term universality refers to phenomena
which are a result of a two-body scattering length $a$ much
larger than the range $R$ of the underlying potential and do not
depend on any further parameters describing the two-body interaction.
The non-relativistic three-body system also exhibits
universal properties if $a\gg R$, but an additional
three-body parameter is needed for the theoretical description of
observables. Therefore, one three-body
observable can be used (e.g. the minimum of the
three-body recombination rate $a_{*0}$)
to predict all other low-energy observables of such systems.
A particularly interesting signature of universality
in the three-body system is a tower of infinitely many bound states
({\it Efimov states}) in the limit $a = \pm \infty$
with an accumulation point at the 
scattering threshold and a geometric spectrum : 
\begin{eqnarray}
E^{(n)}_T = (e^{-2\pi/s_0})^{n-n_*} \hbar^2 \kappa^2_* /m,
\label{kappa-star}
\end{eqnarray}
where $\kappa_*$ is the binding wavenumber of the branch 
of Efimov states labeled by $n_*$. The three-body system 
displays therefore discrete scaling symmetry in the universal
limit with a scaling factor  factor $e^{\pi/s_0}$.
In the case of identical bosons, $s_0 \approx 1.00624$
and the discrete scaling factor is $e^{\pi/s_0} \approx 22.7$.
These results were first derived in the 1970's by Vitaly Efimov \cite{Efimov70,Efimov71} 
and were rederived in the last decade in the framework of effective field
theories (EFT) \cite{Bedaque:1998kg,Braaten:2004rn}.

Recently, experimental evidence for Efimov physics was found by 
the Innsbruck group \cite{Grimm06}. Using a magnetic field to control the
scattering length {\it via} a Feshbach resonance, they
measured the recombination rate of cold $^{133}$Cs atoms and
observed a resonant enhancement in the three-body recombination
rate at $a\approx -850 a_0$ which occurs because 
an Efimov state is close to the 3-atom threshold for that value of $a$.
The three-body recombination rate for atoms with large scattering length
at non-zero temperature has been calculated with a number of different
models or based on the universality of atoms with large scattering lengths
\cite{DSE-04,LKJ07,Jonsell06,YFT06,MS07}.
However, a striking way to demonstrate universality is to describe observables 
of one system with information which has been extracted
from a completely different system.
In \cite{Braaten:2006qx}, the authors considered Efimov's
radial laws which parameterize the three-atom S-matrix in
terms of six real universal functions which depend
only on a dimensionless scaling variable, $x=(m a^2 E/\hbar^2)^{1/2}$, and phase factors which
only contain the three-body parameter. In this work, 
simplifying assumptions justified over a restricted range of $x$ were 
made to reduce the six universal functions required to parameterize
the three-body recombination rate to just a single function.  This function was 
then extracted from microscopic calculations of the recombination rates for
$^4$He atoms by Suno {\it et al.}\cite{Suno:2002}. In a recent paper, Shepard
\cite{Shepard:2007gj} calculated the recombination rates from atom-dimer elastic 
scattering phase shifts for four different $^4$He potentials
(the so-called HDFB, TTY, LM2M2 and HFDB3FCII potentials)
and was able to obtain two universal functions.

Here, we relax all but one of the simplifying assumptions
made in \cite{Braaten:2006qx} and extract a set of three independent 
universal functions capable of parameterizing the three-body
recombination rate over a wide range of energies.
We test the performance of these universal functions
using ``data'' generated from phase shift calculations\cite{Shepard:2007gj}
employing artificial short-range $^4$He potentials.
Finally, we use the new universal functions to calculate the scattering length and
temperature dependent recombination rate for $^{133}$Cs atoms as measured
by the Innsbruck group\cite{Grimm06} and comment on our results.

\section{Three-Body Recombination}
Three-body recombination is a process in which three atoms collide
to form a diatomic molecule (dimer). If the scattering length is positive
and large compared to the range of the interaction, we have to 
differentiate between deep and shallow dimers. Shallow dimers have
an approximate binding energy of 
$E_{\rm shallow}\simeq\hbar^2/(m a^2)\ll\hbar^2/(m R^2)$. The binding 
energy of deep dimers cannot be expressed in terms of the effective 
range parameters and
 $E_{\rm deep}\gtrsim\hbar^2/(m R^2)$. If the underlying interaction supports
deep bound states, recombination processes can occur for either sign of $a$. 
In a cold thermal gas of atoms, recombination processes lead to a change in
the number density of atoms $n_A$
\beq
\frac{\hbox{d}}{\hbox{d}t}n_A=-L_3\,n_A^3~,
\eeq
where $L_3$ denotes the loss rate constant.
The recombination coefficient, to which $L_3$ is proportional, 
can be decomposed into
\beq
K_3(E)=K_{\rm shallow}(E)+K_{\rm deep}(E)~,
\eeq
and the recombination rate into the shallow dimer can be further
decomposed into contributions from the channels in which the
the total orbital angular momentum of the three atoms has a
definite quantum number $J$ according to 
\beq
K_{\rm shallow}(E)=\sum_{J=0}^{\infty}K^{(J)}(E)~.
\eeq

For now, let us consider recombination {\it via} the shallow dimer only.  
If the collision energy $E$ is small compared to the
natural energy scale $\hbar^2/(m R^2)$, the recombination
rate $K_{\rm shallow}(E)$ is a universal function of the
collision energy $E$ , scattering length $a$ and three-body
parameter $a_{*0}$. 
The universal function depends on the dimensionless scaling variable 
defined as
\beq
x=(m a^2E/\hbar^2)^{1/2}~.
\eeq
For $J>0$ the recombination rate does not depend on the
three-body parameter  $a_{*0}$ and the implications of
universality are therefore particularly simple, namely 
\beq
\label{eq:KJ}
K^{(J)}=f_J(x) \hbar a^4/m~.
\eeq
However, $K^{(0)}$ depends log-periodically on $a_{*0}$ ({\it this} is
the signature of Efimov physics!) and is
related to the S-matrix for elastic atom-dimer scattering through
\beq
K^{(0)}(E)=\frac{k}{x^4}(1-|S_{AD,AD}|^2)~,
\eeq
Efimov's radial law then gives the dependence on complex
{\it universal} functions and the three-body parameter $a_{*0}$
which defines the scattering length for which the
recombination rate has a minimum as
\beq
\label{eq:smatrix}
S_{AD,AD}=s_{22}(x)+\frac{s_{12}^2(x)\,e^{2is_0\ln(a/a_{*0})}}
{1-s_{11}(x)e^{2is_0\ln(a/a_{*0})}}~.
\eeq
The functions $s_{11}$ and $s_{12}$ are known at threshold
\begin{eqnarray}
\label{eq:threshold}
\nonumber
s_{11}(0)&=&-e^{-2\pi s_0}~,\\
\nonumber
s_{12}(0)&=&\sqrt{1-e^{-4\pi s_0}}e^{i\delta_\infty}~,\\
s_{22}(0)&=&e^{2i\delta_\infty}e^{-2\pi s_0}~,
\end{eqnarray}
with $\delta_\infty=1.737$. It follows that $|s_{11}(0)|\simeq 0.002$.
The first simplifying assumptions being made in \cite{Braaten:2006qx}
was that this function remains small ({\it i.e.}; $\ll 1$) for all $x$ and can 
be ignored. Then the energy dependent recombination rate can be written as
\bea
\label{eq:recrate1}
\nonumber
K^{(0)}(E)&=&\frac{144 \sqrt{3}\pi^2}{x^4}\Bigl[1-\bigl(r_{22}^2-r_{12}^4
+2 r_{22}r_{12}^2 \cos[\Phi
+2 s_0 \log(a/a_{*0})]\bigr)\Bigr]\frac{\hbar a^4}{m}~,\\
\eea
where we have set $s_{ij}=r_{ij}\exp(i\phi_{ij})$
and $\Phi=\phi_{22}-2\phi_{12}$.
Under the assumption that $s_{11}$ can be neglected the recombination
rate depends therefore on the three real-valued function
 $r_{12}(x)$, $r_{22}(x)$ and $\Phi(x)$.
It is worth noting that the expression in Eq.~(\ref{eq:recrate1})
is symmetric under exchange of $r_{12}^2$ and $r_{22}$. However,
the threshold conditions in Eq.~(\ref{eq:threshold}) can be used to 
to attribute the correct fit solutions to the universal function.

As also discussed in Ref.~\cite{Braaten:2006qx}, the effects of
deep dimers can easily be incorporated through one
additional parameter $\eta_*$ by making the substitution
\beq
\ln a_{*0} \rightarrow \ln a_{*0} - i\eta_*/s_0~
\eeq 
in, {\it e.g.} Eq.~(\ref{eq:recrate1}). Employing unitarity the resulting
effect on the recombination into shallow dimers can be
written as \cite{Braaten:2008kx}
\bea
\nonumber
K_{\rm shallow}^{(0)}(E) &=& \frac{144 \sqrt{3} \pi^2}{x^4}
\biggl( 1 
- \left| s_{22}(x) + s_{12}(x)^2 e^{2i \theta_{*0} -2 \eta_*}\right|^2
- (1 - e^{-4 \eta_*}) |s_{12}(x)|^2 \biggr) 
\frac{\hbar a^4}{m} \,.\\
\label{Kshallow-s}
\eea
Note that in deriving this expression we assumed again
that $s_{11}\approx 0$.
In the same manner one can derive an expression for the recombination rate
into deep dimers 
\begin{eqnarray}
K_{\rm deep}(E) &=& 
\frac{144 \sqrt{3}\pi^2}{ x^4}(1 - e^{-4 \eta_*})
\big( 1 - |s_{12}(x)|^2 \big) 
\frac{\hbar a^4}{m} \,.
\label{Kdeep-s}
\end{eqnarray}

\section{Alternative Parameterizations}
\begin{figure}
  \centering
  \includegraphics[width=10cm,angle=0,clip=true]{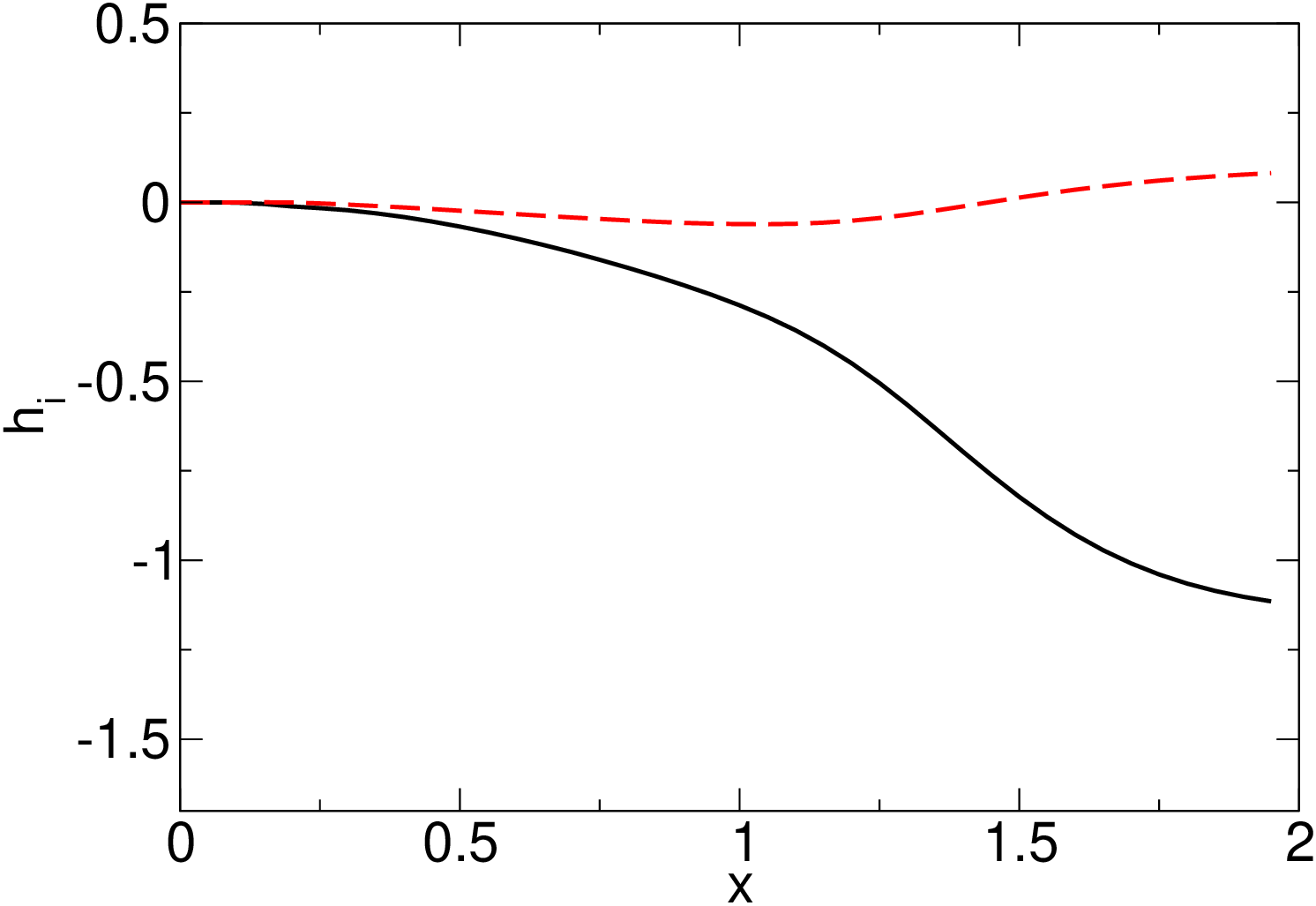}
\caption{The scaling functions $h_1$ (solid line) and $h_2$ as
a function of x.}
\label{fig:hfunctions}
\end{figure}

Starting with S-matrix element for 3-atom to dimer-atom scattering,
it was shown in \cite{Braaten:2006qx} that under the assumption 
$s_{11}=0$ the recombination rate can be written as
\begin{eqnarray}
\nonumber
  \label{eq:reco-h}
K^{(0)}(E)&=&C_{\rm{max}}
\biggl|\biggl(\sin[s_0 \ln(\frac{a}{a_{*0}})]\bigl(1+h_1(x)+i h_3(x)\bigr)\\
&&\hspace{50mm}+\cos[s_0\ln(\frac{a}{a_{*0}})]
\bigl(h_2(x)+i h_4(x)\bigr)\biggr)\biggr|^2\frac{\hbar a^4}{m}~,
\end{eqnarray}
where $C_{\rm{max}}\approx 67.1$ and the $h_i$ are real-valued
functions of $x$. Additionally,it was assumed that the imaginary
part of the above amplitude can be neglected
\begin{eqnarray}
K^{(0)}(E) &=& C_{\rm max} 
\big| \sin[s_0 \ln(a/a_{*0})] (1 + h_1(x)) 
        + \cos[s_0 \ln(a/a_{*0})] h_2(x) \big|^2  \hbar a^4/m\,.
\label{K3-app}
\end{eqnarray}
This is well justified by direct calculations of the $J=0$ 
recombination rates for $^4$He atoms which display pronounced
minima at approximately $E_{\rm breakup}\simeq20$~mK \cite{Suno:2002}
and which can be explained by this assumption. Then the functions
$h_3$ and $h_4$ can be set to 0 in Eq.~(\ref{eq:recrate1}).
The resulting expressions were employed in \cite{Shepard:2007gj}
to extract $h_1$ and $h_2$ for $x<1.1$.
Although $h_1$ and $h_2$ were determined by fitting to values of
$K^{(0)}(E)$ calculated using just two of the four 
atom-atom potentials considered, they were found to accurately
account for the results for all 4 potentials as expected from universality.
We have recalculated the $h$-functions using the results for the
three-body recombination obtained using the LM2M2 and HFDB3FCII
potentials and have fitted a polynomial to our results over the
energy range $0<x<1.2$
\begin{eqnarray}
\label{eq:hfit}
\nonumber
h_1(x)&=&-0.0234437 x + 0.0550298 x^2 - 1.03776 x^3 + 1.18985 x^4 - 
 0.471592 x^5~,\\
h_2(x)&=&0.0338266 x - 0.233836 x^2 + 0.182564 x^3 - 0.0895055 x^4 + 
 0.0461793 x^5~.
\end{eqnarray}
The functions are displayed in Fig.~\ref{fig:hfunctions}.
The effect of deep dimers on the recombination rate into the shallow
dimer can easily be incorporated by making the substitution
$\ln a_{*0} \rightarrow \ln a_{*0} - i\eta_*/s_0$ 
in Eq.~(\ref{eq:reco-h})
\begin{eqnarray}
K^{(0)}(E) &=& C_{\rm max} \Big[ 
\cosh^2 \eta_* \big( \sin[s_0 \ln(a/a_{*0})] (1 + h_1(x)) 
        + \cos[s_0 \ln(a/a_{*0})] h_2(x) \big)^2
\nonumber
\\
&& 
\hspace{0.25cm}
+ \sinh^2 \eta_* \big( \cos[s_0 \ln(a/a_{*0})] (1 + h_1(x)) 
        - \sin[s_0 \ln(a/a_{*0})] h_2(x) \big)^2 \Big] 
\frac{\hbar a^4}{m} \,.
\label{K3-app:deep}
\end{eqnarray}
To take the effects of the recombination
rate into deep dimers into account it was assumed in \cite{Braaten:2006qx}
that $K_{\rm deep}(E)$ is a function varying slowly with energy and that
it can therefore be approximated with
\begin{equation}
  \label{eq:kdeep-h}
  K_{\rm deep}=\frac{C_{\rm}}{4}(1-e^{-4\eta_*})\frac{\hbar a^4}{m}~.
\end{equation}
\section{Extraction of the Universal Functions}
\begin{figure}[t]
\centerline{\includegraphics[width=10cm,angle=0,clip=true]{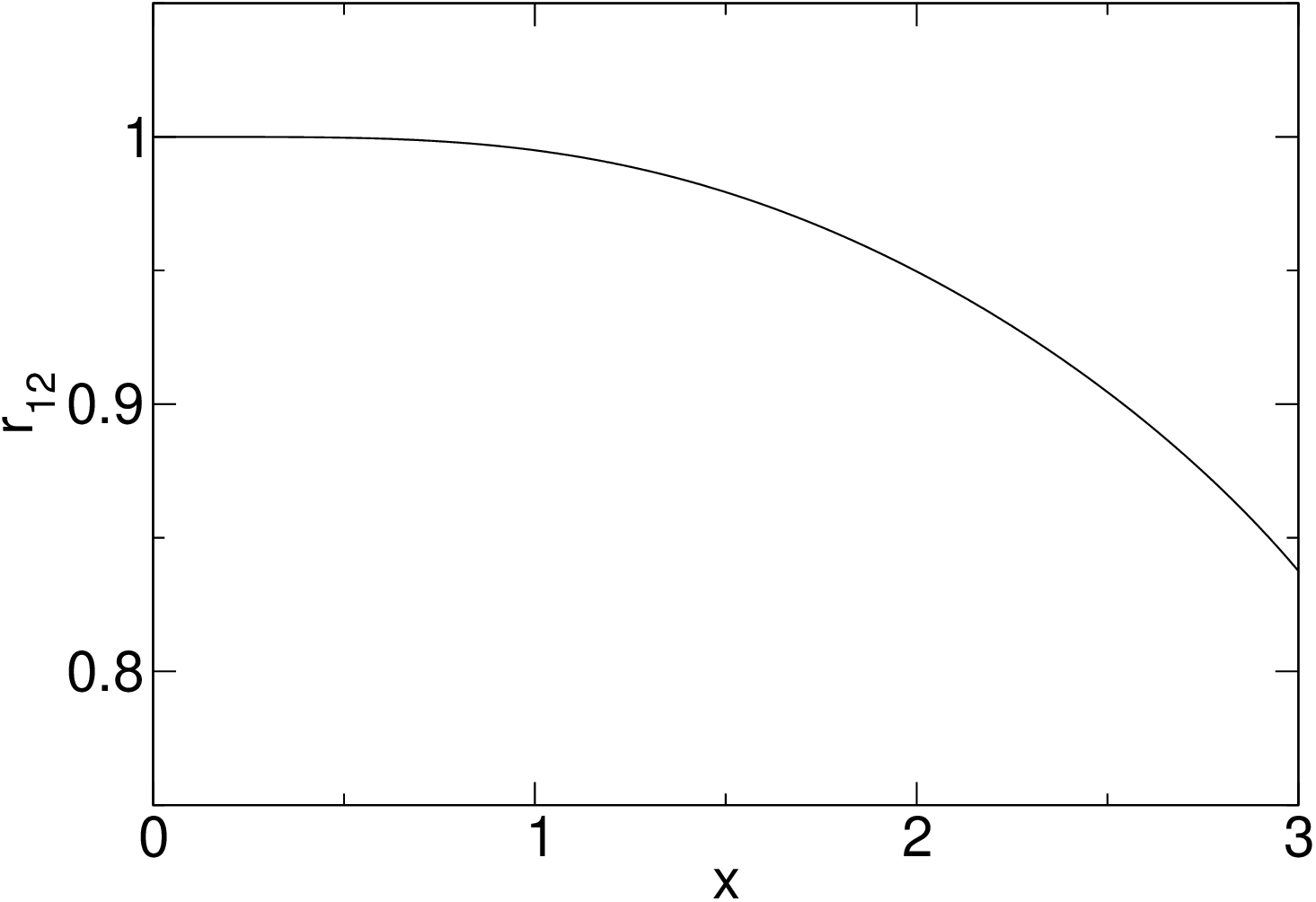}}
\centerline{\includegraphics[width=10cm,angle=0,clip=true]{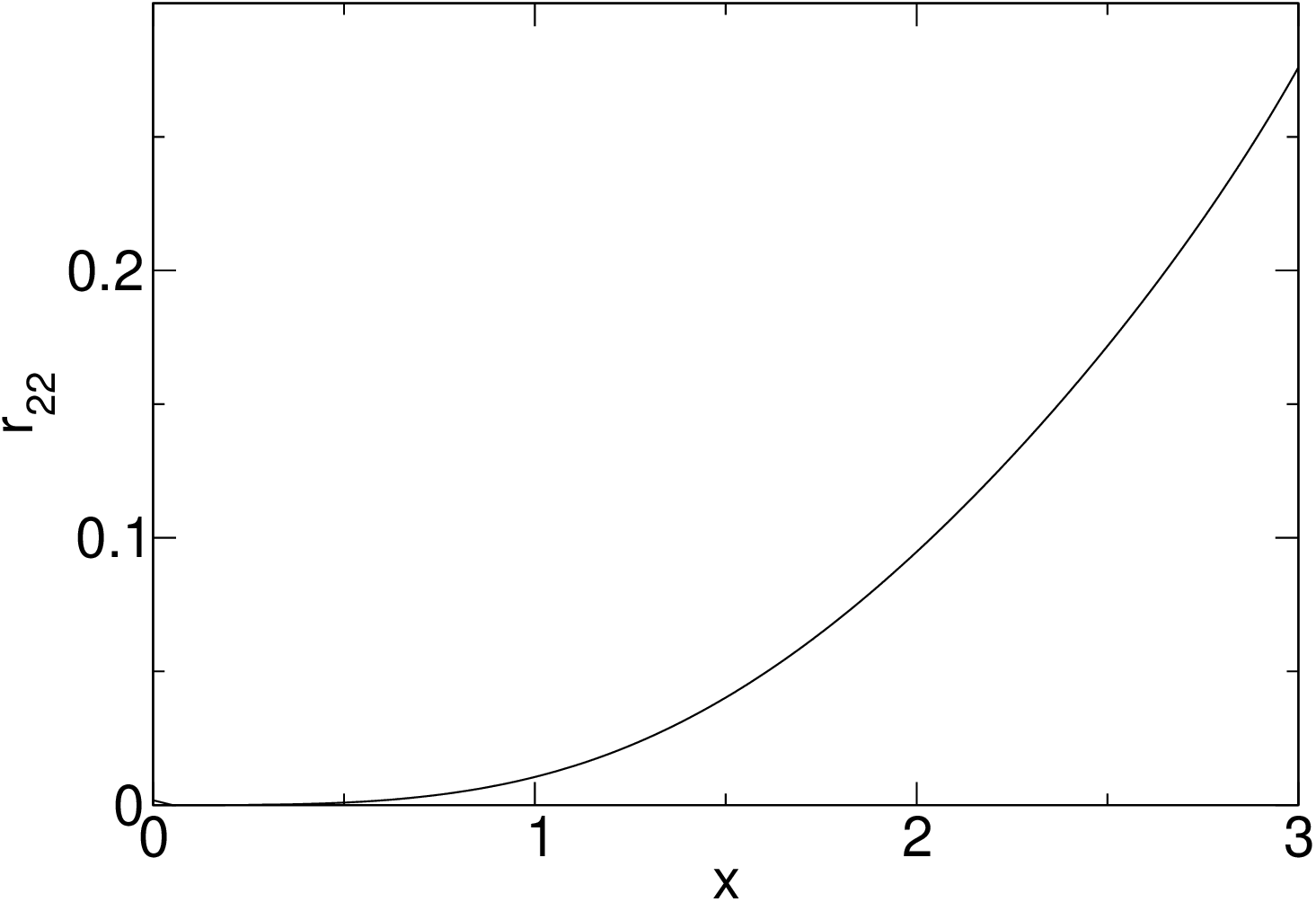}}
\centerline{\includegraphics[width=10cm,angle=0,clip=true]{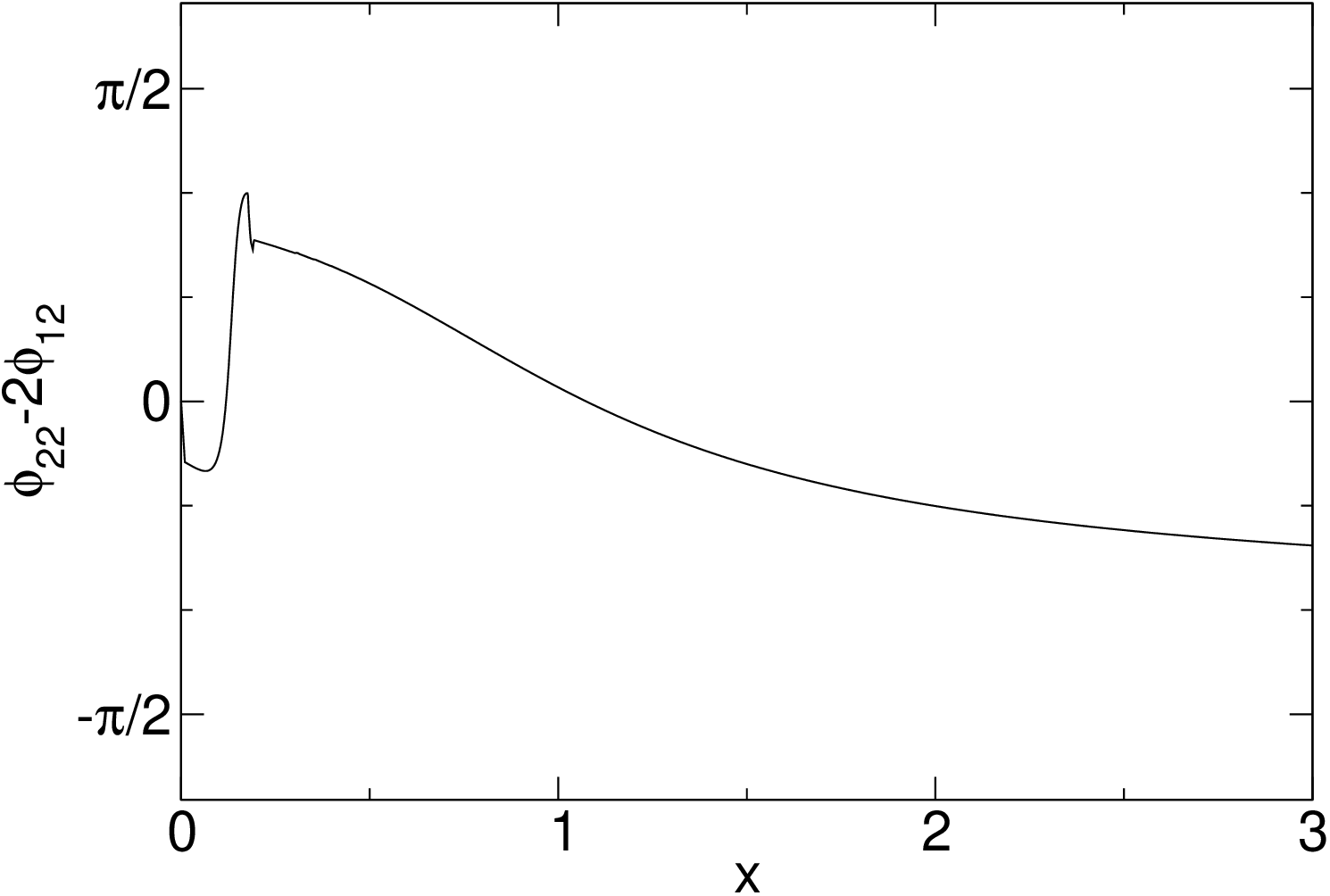}}
\caption{The universal functions $r_{12}, r_{22}$ and $\Phi$
  as function of $x$.}
\label{fig:h1h2}
\end{figure}

By fitting Eq.(\ref{eq:recrate1}) to the recombination rates of all
four $^4$He potentials, we were able to determine the
functions $r_{12}(x)$, $r_{22}(x)$ and $\Phi(x)$. Our results are
smooth functions for $x>0.2$ and the radial functions approach
the known threshold values from Eq.~(\ref{eq:threshold}) for
decreasing $x$. For $x<0.2$, we are not able to find a reliable fit
which is indicated by the rapid variation of the function $\Phi$ in
Fig.~\ref{fig:h1h2} in this region.

To display the qualities of our fit we compare the exact recombination
rates obtained with the TTY and HDFB potentials to the rates calculated
with the newly obtained universal functions
\begin{figure}
  \centering
  \includegraphics[width=10cm,angle=0,clip=true]{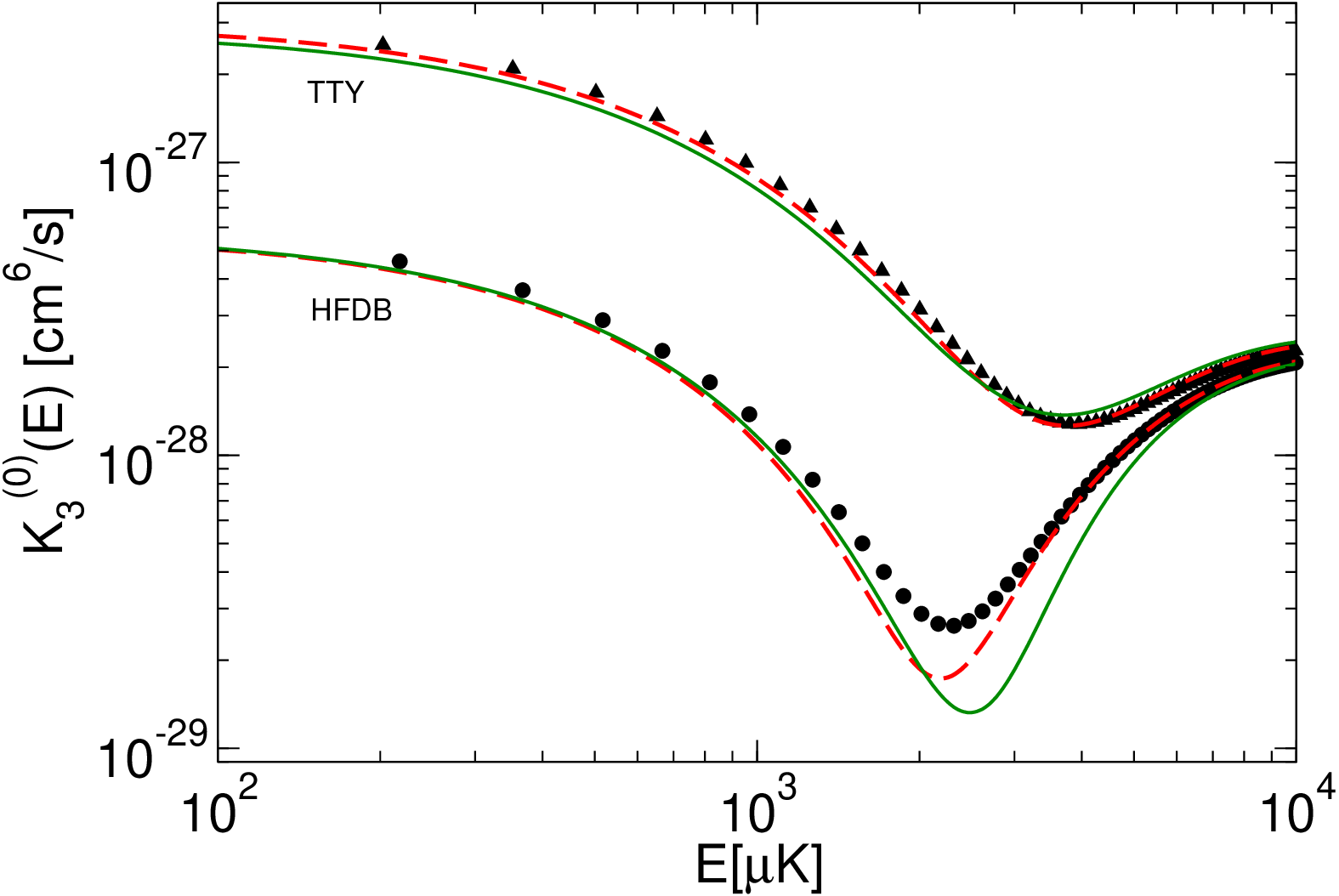}
\caption{The exact recombination rates and the corresponding results
obtained with scaling (solid lines) and universal functions (dashed lines)
of the HFDB (circles) and TTY (triangles) potentials.}
\label{fig:reco-original}
\end{figure}
These results for these functions are displayed in
Fig.~\ref{fig:reco-original}. This figure contains also the 
recombination rate obtained with the $h$-functions. 
While the new set of universal functions seem to provide slightly better
results for the HFDB potential at larger energies, the $h$-functions
perform equally well for these potentials at lower energies.

\begin{figure}[t]
\centerline{\includegraphics[width=10cm,angle=0,clip=true]{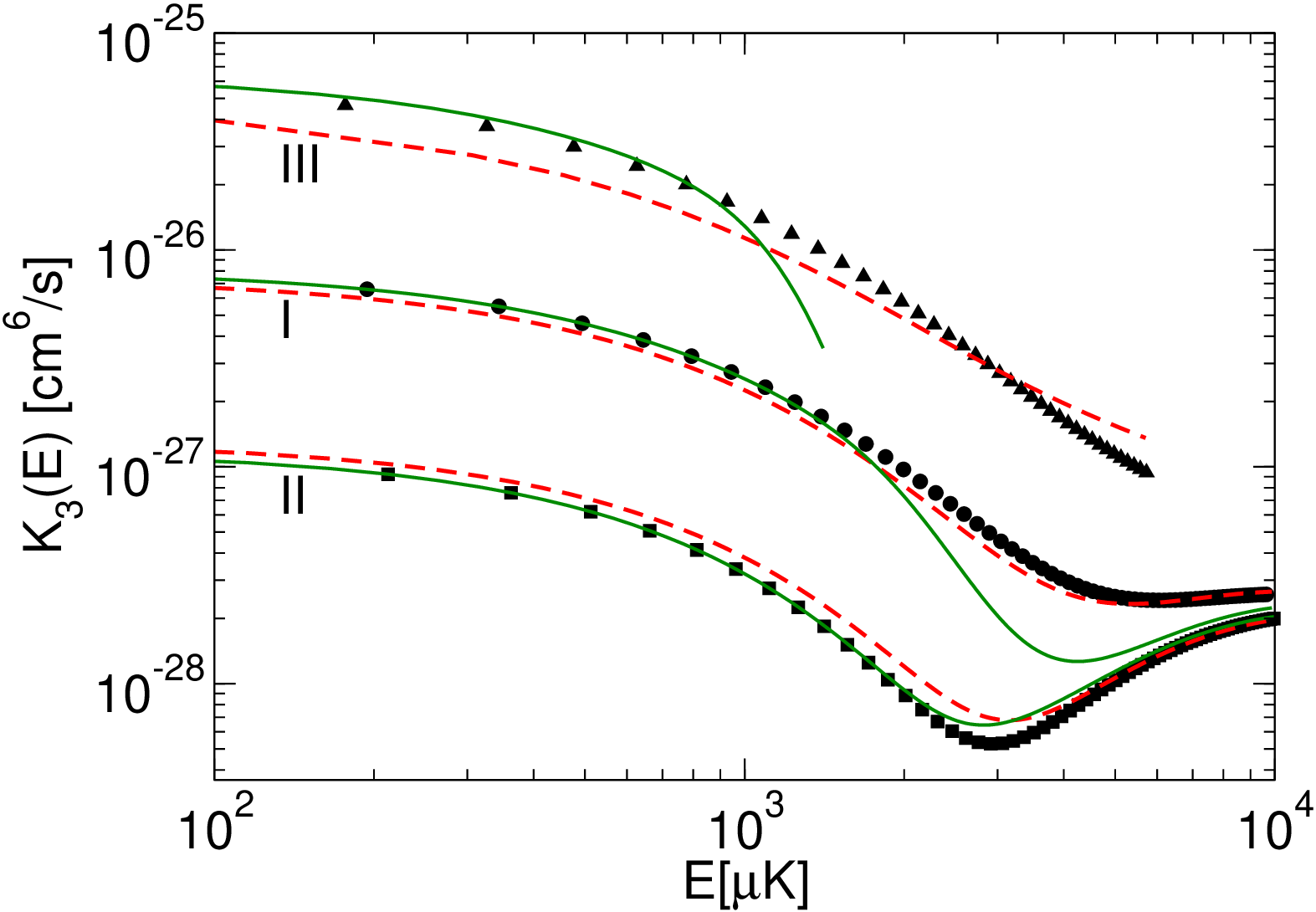}}
\caption{The exact recombination rates and the corresponding results
obtained with the $h$-functions (solid lines) and the
universal functions (dashed lines) for the potentials I (circles),
II (squares) and III (triangles) potentials.}
\label{fig:faux}
\end{figure}
To test our new parameterizations we have generated three artificial
potentials (which we call I, II and III) characterized by 
different three-body parameters $a_{*0}$ (with $a/a_{*0}=1.384, 1.188$ and
$1.780$, respectively) but having approximately
the same ratio of $R/a$ as the real $^4$He potentials used in this work.
We have calculated the recombination rates for these potentials and
use these results to benchmark our universal functions. Our results
are displayed in Fig.\ref{fig:faux}.
We find that the new set of functions is capable of describing
the recombination rates of these potentials over a relatively large range
of $x$. Again we benchmark also the rates obtained with $h_1$ and $h_2$
and find that this set of scaling functions describes the exact results
better at $x<1$ than the scaling functions $r_{12}(x)$, $r_{22}(x)$
and $\Phi(x)$. This is surprising at first sight since one 
certainly expects to obtain a better description of the
recombination rate with three instead of two functions. We speculate
that the functional form in Eq.~(\ref{eq:recrate1}) results in
stronger constrains on the universal functions than
Eq.~(\ref{eq:reco-h}) does on the $h$ functions. All the potentials,
however, contain finite range effects which are not accounted for
in Eq.~(\ref{eq:smatrix}). It is therefore very likely that better
fits -- using the same approximation --
can be obtained from recombination rates calculated in the exact
zero-range limit.
\section{Results for Cesium}
In the previous section we found that we can obtain a very good overall
description of the recombination rate of systems with a large scattering
length if we employ the functions $h_1$ and $h_2$ for energies smaller
than $E_{\rm shallow}$ and the universal function $r_{12}(x)$, $r_{22}(x)$
and $\Phi(x)$ for energies larger than $E_{\rm shallow}$. Using these
functions at energies close to the minimum in the recombination guarantees
a more appropriate treatment of the effect of deep dimers on the recombination
rate, which are expected to have the largest effect in this region.

The form of the functions $f_J(x)$ in Eq.(\ref{eq:KJ}) and therefore the
contribution to the recombination from channels
with higher total angular momentum $J$ has been previously analyzed in
\cite{Braaten:2006qx,Shepard:2007gj}, we thus take these channels into
account by using appropriate parameterizations for the functions $f_J(x)$. 
$^{133}$Cs atoms can recombine into deep and shallow dimers.
As mentioned above, a deep dimer is so strongly bound that it cannot be
described within the EFT for short-range interactions as
the binding energy is larger than $\hbar^2/(m R^2)$. We account for such 
processes by letting $\ln a_{*0} \rightarrow \ln a_{*0} - i\eta_*/s_0$
as also discussed above. 
We then calculate the temperature dependent recombination rate
by calculating
\begin{eqnarray}
\alpha(T) = 
\frac{\int_0^\infty dE \, E^2 \, e^{-E/(k_B T)} \, K_3(E)}
    {6 \int_0^\infty dE \, E^2 \, e^{-E/(k_B T)}} \,.
\label{alpha-T}  
\end{eqnarray}
The weight factor $E^2$ comes from using hyperspherical variables 
for the Jacobi momenta.  

\begin{figure}[t]
\centerline{\includegraphics*[width=10cm,angle=0,clip=true]{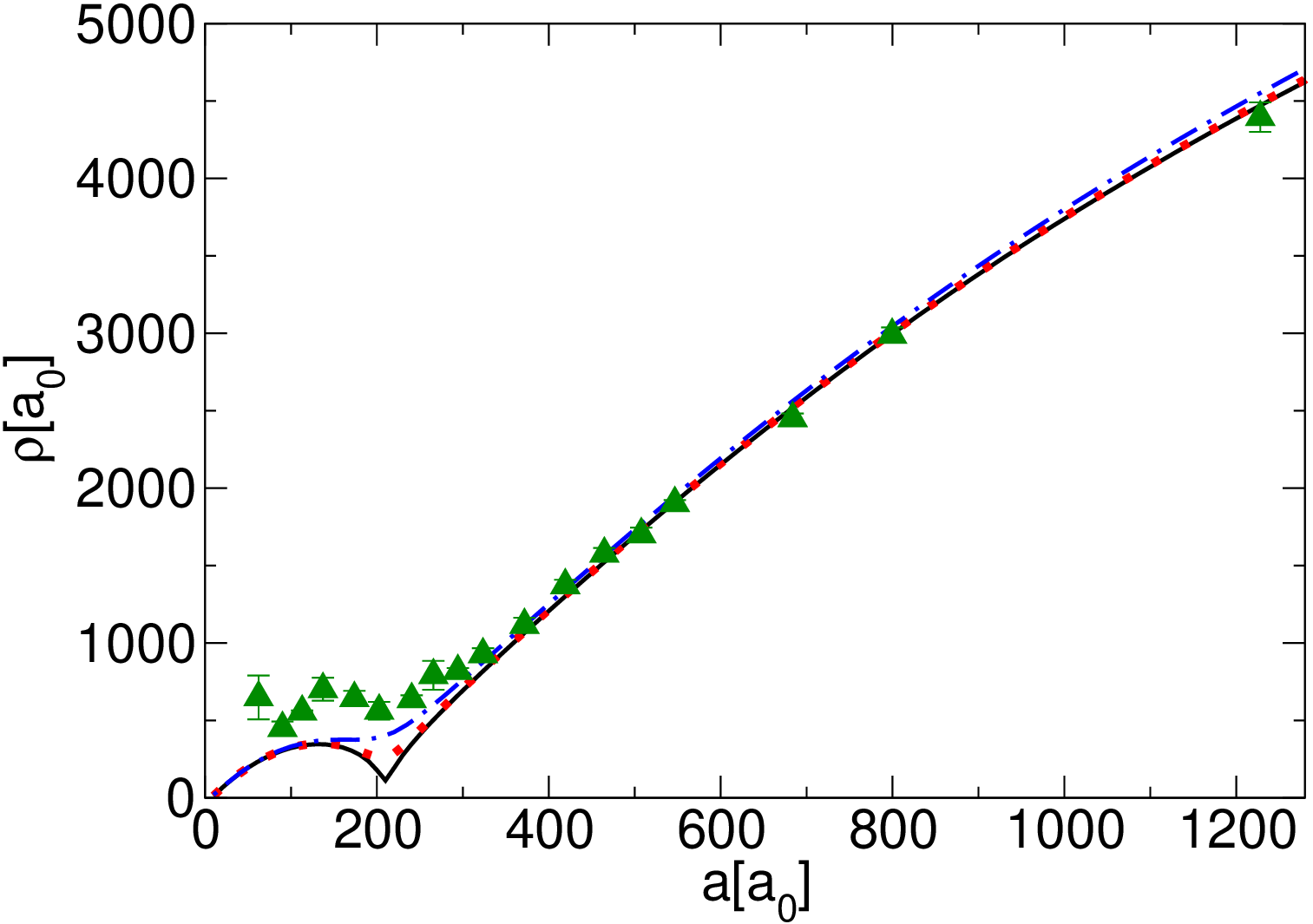}}
\caption{The 3-body recombination length $\rho_3$ for $^{133}$Cs
for $a_{*0}=210 a_0$ and three different values of the parameter
$\eta_*$:0 (solid line), 0.01 (dashed line), and 0.06 (dotted lines)
plotted together with the experimental results of the Innsbruck
experiment (triangles) \cite{Grimm06}}.
\label{fig:rho_cesium}
\end{figure}

In Fig. \ref{fig:rho_cesium} we display our results for the 
recombination length 
$\rho_3 = \left({\textstyle \frac{m K_3}{\sqrt{3} \hbar}} \right)^{1/4}$
of $^{133}$Cs atoms. 
It can be seen that the results agree very well with the
experimental results obtained by the Innsbruck group at $T=200$ nK. 
\section{Summary}
In this paper we have used the results from different $^4$He atom-atom
potentials to extract and to test the predictive power of
universal functions. In doing so, we have relaxed all but one
simplifying assumptions which was made in previous
work \cite{Shepard:2007gj,Braaten:2006qx}.
We have determined a third universal scaling function 
which allows for a description of the three-body recombination 
rate of systems with large scattering length over a greater range of breakup
energies.

We have tested the quality of our parameterizations with
artificial finite range potentials which are 
appreciably different from the original Helium potentials but which 
display universal effects  in three-body sector.
We have found that our three real universal functions can describe
the recombination of these
artificial potentials reasonably well which gives further evidence 
that the assumptions made in \cite{Braaten:2006qx} were well justified. 
We also found, however, that the previously calculated scaling functions
$h_1$ and $h_2$ give an overall better description of the recombination
rate for energies $E < E_{\rm shallow}$. The scaling functions $h_1$ and $h_2$
which can be represented analytically with a simple polynomial fit
given in Eq.~(\ref{eq:hfit}) are therefore a useful tool to test recombination
rate calculations for systems with large scattering length. 

Finally, we have used both sets of universal functions together
to compute the recombination length for $^{133}$Cs atoms for different
values of the parameter $\eta_*$ which approximately accounts for the
effect of deep dimer states and have compared  our results with experimental
data obtained by the Innsbruck group\cite{Grimm06}. 

Although our results show very good agreement with the data, sensitivity to 
$\eta_*$ is insufficient to permit a precise determination of this parameter.
Overall, we consider our results to be an excellent example of
how few-body systems with large scattering length exhibit
universal features. The low-energy properties of $^4$He atoms
allow us to compute accurately the low-energy properties of a gas of a
completely 
different element, $^{133}$Cs, which at first glance has little in common with $^4$He.
Nevertheless, we point out that the results cannot be thought of as 
complete treatment of the problem at hand. For example, not only did we 
make the assumption that $s_{11}$ does not contribute significantly to the 
recombination coefficients, we also extracted the functions from data sets obtained with
finite range potentials. Although the impact of range corrections
is known to be small for realistic Helium atom-atom potentials as $R/a\sim 0.1$,
it needs to be pointed out that range corrections are expected to be
sizable for large enough energies. To obtain all universal functions
$s_{ij}$ relevant to the recombination rate, a calculation in the
limit $R\rightarrow 0$ seems therefore to be necessary
\footnote{This has been done \cite{Braaten:2008kx} in which appeared after
the first submission of this paper}.
Furthermore, it is already understood how to include range corrections
systematically in the framework of effective field theory
\cite{Bedaque:2002yg,Hammer:2001gh,Platter:2006ev}.
Indeed, this approach has already been used to calculate range corrections to 
the recombination rate into a shallow dimer \cite{Hammer:2006zs,Platter:2008cx}. Thus, further
effort should be devoted to include these effects in the calculation
of the energy-dependent recombination rate.
\begin{acknowledgments}
We are thankful to Eric Braaten and Daniel Phillips for useful discussions and
comments on the manuscript.
This work was supported in part by the Department of Energy under
grant DE-FG02-93ER40756, by the National Science
Foundation under Grant No.~PHY--0354916.
\end{acknowledgments}

 

\begin{thebibliography}{199} 

\bibitem{Efimov70}
V.~Efimov,
Phys.\ Lett.\ {\bf 33B}, 563 (1970).

\bibitem{Efimov71}
V.~N.~Efimov, 
Sov.\ J.\ Nucl.\ Phys.\ {\bf 12}, 589 (1971).

\bibitem{Bedaque:1998kg}
  P.~F.~Bedaque, H.-W.~Hammer and U.~van Kolck,
  Phys.\ Rev.\ Lett.\  {\bf 82}, 463 (1999).

\bibitem{Braaten:2004rn}
  E.~Braaten and H.-W.~Hammer,
  Phys.\ Rept.\  {\bf 428}, 259 (2006).

\bibitem{Grimm06}
T. Kraemer, M.~Mark, P.~Waldburger, J.G.~Danzl, C.~Chin, B.~Engeser, 
A.D.~Lange, K.~Pilch, A.~Jaakkola, H.-C.~N\"agerl, and R.~Grimm, 
Nature {\bf 440}, 315 (2006).

\bibitem{DSE-04}
J.P.~D'Incao, H.~Suno, and B.~D.~Esry,
Phys.\ Rev.\ Lett.\ {\bf 93}, 123201 (2004).

\bibitem{LKJ07}
M.D.~Lee, T.~Koehler and P.S.~Julienne,
Phys.\ Rev.\ A {\bf 76}, 012720 (2007).

\bibitem{Jonsell06}
S.~Jonsell,
Europhys.\ Lett.\ {\bf 76}, 8 (2006).

\bibitem{YFT06}
M.T.~Yamashita, T.~Frederico, and L.~Tomio,
Phys.\ Lett.\ A {\bf 363}, 468 (2007).

\bibitem{MS07}
P.~Massignan and H.T.C.~Stoof, 
Phys. Rev. A 78, 030701 (2008).

\bibitem{Braaten:2006qx}
  E.~Braaten, D.~Kang and L.~Platter,
  Phys.\ Rev.\  A {\bf 75}, 052714 (2007).

\bibitem{Suno:2002}
H.~Suno, B.~D.~Esry, C~.~H.~ Greene, and J.~P.~Burke,
Phys. Rev. A 65, 042725 (2002).


\bibitem{Shepard:2007gj}
  J.~R.~Shepard,
  Phys.\ Rev.\  A {\bf 75}, 062713 (2007).



\bibitem{Bedaque:2002yg}
  P.~F.~Bedaque, G.~Rupak, H.~W.~Griesshammer and H.~W.~Hammer,
  Nucl.\ Phys.\  A {\bf 714}, 589 (2003).

\bibitem{Hammer:2001gh}
  H.-W.~Hammer and T.~Mehen,
  Phys.\ Lett.\  B {\bf 516}, 353 (2001).

\bibitem{Platter:2006ev}
  L.~Platter and D.~R.~Phillips,
  Few Body Syst.\  {\bf 40}, 35 (2006).

\bibitem{Braaten:2008kx}
  E.~Braaten, H.~W.~Hammer, D.~Kang and L.~Platter,
 Phys.\ Rev.\  A {\bf 78}, 043605 (2008).


\bibitem{Hammer:2006zs}
  H.-W.~Hammer, T.~A.~Lahde and L.~Platter,
  Phys.\ Rev.\  A {\bf 75}, 032715 (2007).


\bibitem{Platter:2008cx}
  L.~Platter, C.~Ji and D.~R.~Phillips,
 Phys.\ Rev.\  A {\bf 79}, 022702 (2009).

\end{thebibliography}
\end{document}